\documentstyle[preprint,aps,floats,psfig,amssymb,axodraw]{revtex}
\catcode`@=11
\def\references{%
\ifpreprintsty
\bigskip\bigskip
\hbox to\hsize{\hss\large \refname\hss}%
\else
\vskip24pt
\hrule width\hsize\relax
\vskip 1.6cm
\fi
\list{\@biblabel{\arabic{enumiv}}}%
{\labelwidth\WidestRefLabelThusFar  \labelsep4pt %
\leftmargin\labelwidth %
\advance\leftmargin\labelsep %
\ifdim\baselinestretch pt>1 pt %
\parsep  4pt\relax %
\else %
\parsep  0pt\relax %
\fi
\itemsep\parsep %
\usecounter{enumiv}%
\let\p@enumiv\@empty
\def\theenumiv{\arabic{enumiv}}%
}%
\let\newblock\relax %
\sloppy\clubpenalty4000\widowpenalty4000
\sfcode`\.=1000\relax
\ifpreprintsty\else\small\fi
}
\catcode`@=12
\begin{document}
\def\beq{\begin{equation}}
\def\eeq{\end{equation}}
\def\d{\delta}
\def\fourG{{{}^{(4)}G}}
\def\4R{{{}^{(4)}R}}
\def\H{{\cal H}}
\def\K{{\kappa}}
\def\mh{m_h^{}}
\def\vev#1{{\langle#1\rangle}}
\def\gev{{\rm GeV}}
\def\tev{{\rm TeV}}
\def\fbi{\rm fb^{-1}}
\def\lsim{\mathrel{\raise.3ex\hbox{$<$\kern-.75em\lower1ex\hbox{$\sim$}}}}
\def\gsim{\mathrel{\raise.3ex\hbox{$>$\kern-.75em\lower1ex\hbox{$\sim$}}}}
\newcommand{\hmu}{{\hat\mu}}
\newcommand{\hnu}{{\hat\nu}}
\newcommand{\hrho}{{\hat\rho}}
\newcommand{\hh}{{\hat{h}}}
\newcommand{\hg}{{\hat{g}}}
\newcommand{\hk}{{\hat\kappa}}
\newcommand{\tA}{{\widetilde{A}}}
\newcommand{\tP}{{\widetilde{P}}}
\newcommand{\tF}{{\widetilde{F}}}
\newcommand{\th}{{\widetilde{h}}}
\newcommand{\tp}{{\widetilde\phi}}
\newcommand{\tchi}{{\widetilde\chi}}
\newcommand{\te}{{\widetilde\eta}}
\newcommand{\vn}{{\vec{n}}}
\newcommand{\vm}{{\vec{m}}}
\newcommand{\A}{A}
\newcommand{\mmu}{\mu}
\newcommand{\mnu}{\nu}
\newcommand{\ii}{i}
\newcommand{\jj}{j}
\newcommand{\jl}{[}
\newcommand{\jr}{]}
\newcommand{\ml}{\sharp}
\newcommand{\mr}{\sharp}
\renewcommand{\theenumi}{\roman{enumi}}
\renewcommand{\labelenumi}{(\theenumi)}
\newcommand{\da}{\dot{a}}
\newcommand{\db}{\dot{b}}
\newcommand{\dn}{\dot{n}}
\newcommand{\dda}{\ddot{a}}
\newcommand{\ddb}{\ddot{b}}
\newcommand{\ddn}{\ddot{n}}
\newcommand{\pa}{a^{\prime}}
\newcommand{\pb}{b^{\prime}}
\newcommand{\pn}{n^{\prime}}
\newcommand{\ppa}{a^{\prime \prime}}
\newcommand{\ppb}{b^{\prime \prime}}
\newcommand{\ppn}{n^{\prime \prime}}
\newcommand{\fda}{\frac{\da}{a}}
\newcommand{\fdb}{\frac{\db}{b}}
\newcommand{\fdn}{\frac{\dn}{n}}
\newcommand{\fdda}{\frac{\dda}{a}}
\newcommand{\fddb}{\frac{\ddb}{b}}
\newcommand{\fddn}{\frac{\ddn}{n}}
\newcommand{\fpa}{\frac{\pa}{a}}
\newcommand{\fpb}{\frac{\pb}{b}}
\newcommand{\fpn}{\frac{\pn}{n}}
\newcommand{\fppa}{\frac{\ppa}{a}}
\newcommand{\fppb}{\frac{\ppb}{b}}
\newcommand{\fppn}{\frac{\ppn}{n}}
\newcommand{\matbr}{\mathbb{R}}
\newcommand{\matbh}{\mathbb{H}}
\newcommand{\sfk}{\sf{k}}
\newcommand{\dA}{\dot{A_0}}
\newcommand{\dB}{\dot{B_0}}
\newcommand{\fdA}{\frac{\dA}{A_0}}
\newcommand{\fdB}{\frac{\dB}{B_0}}

\newcommand{ \slashchar }[1]{\setbox0=\hbox{$#1$}   
   \dimen0=\wd0                                     
   \setbox1=\hbox{/} \dimen1=\wd1                   
   \ifdim\dimen0>\dimen1                            
      \rlap{\hbox to \dimen0{\hfil/\hfil}}          
      #1                                            
   \else                                            
      \rlap{\hbox to \dimen1{\hfil$#1$\hfil}}       
      /                                             
   \fi}                                             %

\tighten
\preprint{ \vbox{
\hbox{MADPH--00-1186}
\hbox{hep-th/0007039}}}
\draft
\title{Brane Networks in $AdS$ Space}
\author{Jing Jiang, Tianjun Li and Danny Marfatia\footnote{
 {\it {lastname}}@pheno.physics.wisc.edu }}
\vskip 0.3in
\address{Department of Physics, University of Wisconsin--Madison, WI 53706}
\vskip 0.1in

\maketitle

\begin{abstract}
{\rm 
We present static solutions to Einstein's equations corresponding to the most 
general network of $N$ orthogonal
families of $(2+N)$-branes in $(4+N)$-dimensional $AdS$ spacetimes. The
bulk cosmological constant can take a different value in each cell enclosed by 
intersecting branes and the extra dimensions can be compact or noncompact. 
In each family of branes the inter-brane spacing is arbitrary. The extra
dimensions may be any or all of the manifolds, ${\mathbb{R}}^1$, 
${\mathbb{R}}^1/{\mathbf{Z}}_2\,$, ${\mathbb{S}}^1$ and 
${\mathbb{S}}^1/{\mathbf{Z}}_2$.  Only when the extra dimensions 
are ${\mathbb{R}}^1$ or/and 
${\mathbb{R}}^1/{\mathbf{Z}}_2\,$, can 
networks consisting solely of positive tension branes be constructed. 
Such configurations may find application in models with 
localized gravity and symmetry breaking by shining.
}
\end{abstract}
{{\bf Introduction.}}
By now it is redundant to say that Kaluza-Klein theory has been revived. 
The resurgence in interest began with the realization that the string scale
can be much lower than the Planck scale and even close to the
electroweak scale \cite{ewstring}. 
A low string scale provides new avenues on solving the 
hierarchy problem \cite{add,rs1}. The original argument resides
in the fact that a low string scale ($M_X$) may result in
the apparent size of the Planck scale ($M_{Pl}$) due to the 
existence of a large volume ($R^n$) of compact extra 
dimensions, $ M_{Pl}^2 \sim M_X^{n+2} R^n $ \cite{add}. One may wonder, 
as Randall and Sundrum did, if it is necessary to compactify the extra dimensions 
at all given that the only purpose of this compactification is to reproduce
Newtonian gravity at large distances. Instead, localizing
gravity on a 3-brane in a noncompact $AdS_5$ geometry 
was shown to work just as well  \cite{rs2}.
Taking this idea further, it was demonstrated that it is possible to localize gravity
at the intersection of $N$ mutually intersecting $(2+N)$-branes in a $(4+N)$-dimensional 
spacetime with a bulk cosmological constant \cite{addk}. 
The single 3-brane in which the branes intersect  could well be our universe.
Subsequently, a generalization was made to the case where the slices of spacetime
created by the intersecting branes could have different bulk cosmological constants 
\cite{cs}. The natural next step is to suppose that there are many brane junctions
formed by intersecting branes, and therefore a multitude of 3-branes which could 
in principle be responsible for symmetry breaking by shining \cite{shine}.  
A hexagonal network of supersymmteric branes was constructed in \cite{nam}. 
A solution involving intersecting families of $(2+N)$-branes in $(4+N)$-dimensional $AdS$ spacetimes
was presented in Ref. \cite{kaloper}. However, the bulk cosmological constant was taken
to be the same everywhere, thus making it necessary to include negative tension
branes. With the assumption that the bulk cosmological constant is global, it is not possible to find a solution with only positive tension branes when there
is more than one brane junction.
\\

Our purpose is to construct the most general network of $N$ orthogonal
families of $(2+N)$-branes in $(4+N)$-dimensional $AdS$ spacetimes where the
bulk cosmological constant can take an arbitrary value in each cell enclosed 
by intersecting branes and the extra dimensions can be compact or noncompact. 
It is now
possible to find solutions involving only positive tension branes.  
To simplify notation we 
 illustrate the minimal model that embodies all the salient features of
the general case. We present the equations in a form that facilitates
easy generalization.     
\\

{{\bf Network Construction.}}
Consider a network of four orthogonal families of $6$-branes in $AdS_8$. The four
extra dimensions $y$, $z$, $v$ and $w$ are ${\mathbb{R}}^1$, 
${\mathbb{R}}^1/{\mathbf{Z}}_2\,$,  ${\mathbb{S}}^1$ and 
${\mathbb{S}}^1/{\mathbf{Z}}_2$ and will be denoted by  ${\mathbb{R}}$,  
${\mathbb{H}}$, ${\mathbb{S}}$ and  ${\mathbb{I}}$, respectively. Assume that 
there are three branes along each of the extra dimensions. 
The locations of the branes along the extra dimensions 
$y$, $z$, $v$ and $w$ are labelled by $y_{j}\,,z_{k}\,,v_{l}\,$ and $w_{m}$ 
respectively, and the branes are arranged in the order 
of increasing values of the subscripts. In each family of branes the inter-brane spacing 
is arbitrary. 
In most of what follows, we explicitly display the $N$-dependence
even though we present results for $N=4$. 
The $(3+N)$-dimensional metric on the brane labelled by its position $y_j$ is
\begin{eqnarray}
(g_j^{\mathbb{R}})_{\hat{A}\hat{B}} 
\equiv g_{\hat{A}\hat{B}}
(y = y_j) \quad
\rm{with}\ \ \  \hat{A},\hat{B} \neq 4 ~,
\end{eqnarray}
where $g_{\hat{A}\hat{B}}$ is the $(4+N)$-dimensional metric and 
$A,B=0,\ldots,7$
 (Note that $(x_0,\dots,x_3,y,z,v,w)\equiv (0,\ldots,3,4,5,6,7)$). Similar equations define the metric on the branes at 
$z_{k}\,,v_{l}\,$, and $w_{m}$. We choose the metric signature $(-,+,\dots,+)$. The
$(4+N)$-dimensional gravitational action is
\begin{eqnarray}
S &=& S^0 + S^{\mathbb{R}} + S^{\mathbb{H}} 
+ S^{\mathbb{S}} + S^{\mathbb{I}}~,  \\
S^0 &=& \int d^4x~ dy~ dz~ dv~ dw ~\sqrt{-g} 
~\{ \frac{1}{2\,\kappa^2} R
- \Lambda(y,z,v,w) \} ~, \label{grav}\\
S^{\mathbb{R}} &=& -\sum_{j=0}^{2} \int d^4x ~dy ~dz ~dv ~dw 
~\sqrt{-g_j^{\mathbb{R}}} 
~ V_j^{\mathbb{R}}
~\delta( y - y_j ) ~, \\
S^{\mathbb{H}} &=& -\sum_{k=0}^{2} \int d^4x ~dy ~dz ~dv ~dw 
~\sqrt{-g_k^{\mathbb{H}}}
~ V_k^{\mathbb{H}} 
~\delta( z - z_k ) ~, \\
S^{\mathbb{S}} &=& -\sum_{l=0}^{2} \int d^4x ~dy ~dz ~dv ~dw 
~\sqrt{-g_l^{\mathbb{S}}}
~ V_l^{\mathbb{S}} 
~\delta ( v - v_l ) ~, \\
S^{\mathbb{I}} &=& -\sum_{m=0}^{2} \int d^4x ~dy ~dz ~dv ~dw 
~\sqrt{-g_m^{\mathbb{I}}}
~ V_m^{\mathbb{I}}  
~\delta ( w - w_m ) ~,
\end{eqnarray}
where 
$\kappa^2= M_X^{-(2+N)}$ is the $(4+N)$-dimensional coupling constant of gravity, 
$M_X$ is the Planck scale in $(4+N)$ dimensions, $R$ is the curvature scalar and
the brane tensions are denoted by $V$ with the superscript representing the type of extra dimension and the subscript indicating the position of the brane. 
$\Lambda(y,z,v,w)$ is the bulk cosmological constant which can have an 
arbitrary value in each cell formed by the intersecting branes. It
can be written as
\begin{eqnarray}
\Lambda(y,z,v,w) &=& \sum_{j=0}^3 \sum_{k=1}^3 \sum_{l=0}^2 \sum_{m=1}^2 
\Lambda_{(j,k,l,m)} ~[\theta(y-y_{j-1})-\theta(y-y_{j}) ]~[\theta(z-z_{k-1})-\theta(z
-z_{k}) ]
 \nonumber \\ &&
\ \ \ \ \ \ \ \ \ \ \ \ \ \ \ \ \ \ \  \ \ \ \ \ \, \times ~[\theta(v-v_{l-1})-
\theta(v-v_{l}) ]~
[\theta(w-w_{m-1})-\theta(w-w_{m}) ] ~,
\end{eqnarray}
where $\Lambda_{(j,k,l,m)}$ is the bulk cosmological constant in the cell defined by $y_{j-1}<y<y_j\,$, \mbox{$z_{k-1}<z<z_k\,$} , $v_{l-1}<v<v_l\,$ and $w_{m-1}<w<w_m\,$. 
 We have set $y_{-1}=-\infty$ and $y_3=z_3=\infty$ for the two noncompact 
extra dimensions. For 
the extra dimension compactified on a circle $\mathbb{S}$, we assume that
$v \in [0, 2 \pi r^{\mathbb{S}}]$ with  $v_{-1}=0$ and $v_2=2 \pi r^{\mathbb{S}}$. The fixed points of the orbifold
extra dimension $\mathbb{I}$, are located at $w_0=0$ and $w_2=\pi  r^{\mathbb{I}}$.

The $(4+N)$-dimensional Einstein equations arising from the above action are
\begin{eqnarray}
G_{AB}&\equiv& R_{AB}-{1 \over 2 } g_{AB} R=  \kappa^2~T_{AB}  =  \kappa^2~(T^0 + T^{\mathbb{R}} + T^{\mathbb{H}} 
+T^{\mathbb{S}} + T^{\mathbb{I}})_{AB} ~, 
\label{Einstein} \\
T^0_{AB} &=& - \Lambda(y,z,v,w) ~g_{AB} ~, \nonumber \\
T^{\mathbb{R}}_{AB} &=& - \sum_{j=0}^2
V_j^{\mathbb{R}} ~\sqrt{\frac{g_j^{\mathbb{R}}}{g}} 
 ~(g_j^{\mathbb{R}})_{\hat{A}\hat{B}}
~\delta_A^{\hat A } ~\delta_B^{\hat B} ~\delta(y-y_j) ~,\nonumber \\
T^{\mathbb{H}}_{AB} &=& - \sum_{k=0}^2
V_k^{\mathbb{H}} ~\sqrt{\frac{g_k^{\mathbb{H}}}{g}} 
 ~(g_k^{\mathbb{H}})_{\hat{A}\hat{B}}
~\delta_A^{\hat A } ~\delta_B^{\hat B} ~\delta(z-z_k) ~,\nonumber \\
T^{\mathbb{S}}_{AB} &=& - \sum_{l=0}^2
V_l^{\mathbb{S}} ~\sqrt{\frac{g_l^{\mathbb{S}}}{g}} 
 ~(g_l^{\mathbb{S}})_{\hat{A}\hat{B}}
~\delta_A^{\hat A } ~\delta_B^{\hat B} ~\delta(v-v_l) ~,\nonumber \\
T^{\mathbb{I}}_{AB} &=& - \sum_{m=0}^2
V_m^{\mathbb{I}} ~\sqrt{\frac{g_m^{\mathbb{I}}}{g}} 
 ~(g_m^{\mathbb{I}})_{\hat{A}\hat{B}}
~\delta_A^{\hat A } ~\delta_B^{\hat B} ~\delta(v-v_m) ~, \nonumber
\end{eqnarray}
where $R_{AB}$ is the  $(4+N)$-dimensional Ricci tensor. Assume the 
metric to be conformally flat and write it as
\begin{eqnarray}
ds_{4+N}^2 & = & \Omega^2
~( \eta_{\mu\nu} ~dx^{\mu} dx^{\nu} + dy^2 + dz^2 + dv^2 + dw^2 ) ~,
\label{metric}
\end{eqnarray}
where $\Omega\equiv\Omega(y,z,v,w)$. The simplest way to proceed is to transform to a conformally related 
spacetime, {\it i.e.} 
\begin{equation}
g_{AB} =\Omega^{2}~ \tilde g_{AB} ~. 
\end{equation}
The Einstein tensor in the two metrics are related by
\begin{equation}
G_{AB} = \tilde G_{AB}+ (2+N)\left[  \tilde{\nabla}_{\! A} \ln \Omega 
 ~\tilde{\nabla}_{\! B} \ln \Omega - \tilde{\nabla}_{\! A} \tilde{\nabla}_{\!B}
\ln \Omega 
+ \tilde g_{AB}  \left(\tilde \nabla^2 \ln \Omega + 
\frac{1+N}{2} ( \tilde{\nabla} \ln \Omega)^2\right)\right]~,
\end{equation}
where the covariant derivatives  $\tilde{\nabla}$ are evaluated with respect to 
the metric $\tilde g$. Since the metric is conformally flat, 
 the covariant derivatives are identical to ordinary derivatives and 
$\tilde G_{AB}=0$. On further simplification, the above equation can be recast to
\begin{eqnarray}
G_{AB} = (2+N)\left[ \Omega \tilde{\nabla}_{\! A} \tilde{\nabla}_{\! B} 
\Omega^{-1}
+  \tilde g_{AB} \left(- \Omega \tilde \nabla^2 \Omega^{-1} + 
\frac{3+N}{2} ~\Omega^2 ( \tilde{\nabla} \Omega^{-1})^2\right)\right],
\end{eqnarray}
where the relation $ \tilde{\nabla}_{\! A}\ln \Omega =-\Omega \tilde{\nabla}_{\! A} \Omega^{-1}\,$ has been employed.
Using the above form of the Einstein tensor, the Einstein equations 
(\ref{Einstein}) can be written as 
\begin{eqnarray}
\frac{\partial^2}{\partial y^2} \Omega^{-1}& =&
\sum_{j=0}^2 \frac{\kappa^2}{(2+N) }~V_j^{\mathbb{R}} ~\delta(y-y_j) ~, 
\label{dd1} \\
\frac{\partial^2}{\partial z^2} \Omega^{-1}& =&
\sum_{k=0}^2 \frac{\kappa^2}{(2+N) }~V_k^{\mathbb{H}} ~\delta(z-z_k) ~, 
\label{dd2} \\
\frac{\partial^2}{\partial v^2} \Omega^{-1}& =&
\sum_{l=0}^2 \frac{\kappa^2}{(2+N) }~V_l^{\mathbb{S}} ~\delta(v-v_l) ~, 
\label{dd3} \\
\frac{\partial^2}{\partial w^2} \Omega^{-1}& =&
\sum_{m=0}^2 \frac{\kappa^2}{(2+N) }~V_m^{\mathbb{I}} ~\delta(w-w_m)  \ \ \ \ \  {\rm{and}}
\label{dd4} \\
(\tilde \nabla \Omega^{-1})^2 &=& 
- \frac{2\, \kappa^2} {(2+N)(3+N)}~ \Lambda(y,z,v,w)\,.
\label{dd0}
\end{eqnarray}

We can relate the fundamental scale $M_X$ to the four-dimensional Planck scale $M_{Pl}$ by
integrating over the extra dimensions in Eq.~(\ref{grav}), 
\begin{eqnarray}
M_{Pl}^2 = M_X^{2+N} \int dy~dz~dv~dw~\Omega^{2+N}~.
\end{eqnarray}
The ``electroweak scale'' $M_{EW}$ (for want of a better term), at each brane junction is 
\begin{eqnarray}
(M_{EW})_{(j,k,l,m)} = M_X 
~\Omega(y = y_j, z = z_k, v = v_l, w = w_m)\,.
\end{eqnarray}
Thus, it is possible to generate hierarchies of scales by choosing $\Omega$ 
appropriately.
\\

{{\bf Network Solutions.}}
We note that $\Omega^{-1}$ is a linear combination of the solutions to 
Eqs. ($\ref{dd1}-\ref{dd4}$). We write
\begin{equation}
\Omega^{-1} = \sigma^{\mathbb{R}}(y) + \sigma^{\mathbb{H}}(z) 
+\sigma^{\mathbb{S}}(v) +\sigma^{\mathbb{I}}(w) + c ~,
\end{equation}
where  $\sigma^{\mathbb{R}}(y)\,, \sigma^{\mathbb{H}}(z) 
\,,\sigma^{\mathbb{S}}(v)$ and $\sigma^{\mathbb{I}}(w)$ satisfy Eqs. 
($\ref{dd1}\,,\ref{dd2}\,,\ref{dd3}$) and ($\ref{dd4}$), respectively.
Again, to simplify notation, define  
\begin{equation}
{\sf{k}}_j^{\mathbb{R}} = \frac{\kappa^2}{2\,(2+N)}~V_j^{\mathbb{R}}\,,
\end{equation} 
with similar definitions for ${\sf{k}}_k^{\mathbb{H}}\,$, ${\sf{k}}_l^{\mathbb{S}}$ and ${\sf{k}}_m^{\mathbb{I}}$. Note that  ${\sf{k}}$ is simply a rescaling of $V$.

Since the evaluation of Eqs. ($\ref{dd1}-\ref{dd4}$) has been carried out in detail in 
Ref.~\cite{li}, we state the solutions without undue elaboration. Along $\mathbb{R}\,$,
\begin{eqnarray}
\sigma^{\mathbb{R}}(y) &=& \sum_{j=0}^2 {\sf{k}}_j^{\mathbb{R}} ~| y - y_j | 
+ {\sf{k}}_c^{\mathbb{R}} ~y ~ {\rm \ \ \ \ \ with \ \ \ }~
\sum_{j=0}^2 {\sf{k}}_j^{\mathbb{R}} > | {\sf{k}}_c^{\mathbb{R}} | ~,
\label{R}
\end{eqnarray}
where ${\sf{k}}_c^{\mathbb{R}}$ is an integration constant. The constraint ensures that $\sigma^{\mathbb{R}}(y)$ is bounded below and unbounded above thus
guaranteeing that  the 
four-dimensional Planck scale is finite. For the semi-infinite interval $\mathbb{H}$,
\begin{eqnarray}
\sigma^{\mathbb{H}}(z) &=& \sum_{k=0}^2 {\sf{k}}_k^{\mathbb{H}} ~| z - z_k | 
+ {\sf{k}}_t^{\mathbb{H}} ~z~ {\rm \ \ \    with \ \ \ }~
{\sf{k}}_t^{\mathbb{H}} = \sum_{k=1}^2 {\sf{k}}_k^{\mathbb{H}}\ \   ~{\rm and}~\ \ 
{\sf{k}}_0^{\mathbb{H}}+ 2\,{\sf{k}}_t^{\mathbb{H}} > 0 ~,
\label{H}
\end{eqnarray}
where the constraint must be satisfied in order to obtain a finite four-dimensional Planck scale. Because of imposing the discrete symmetry $\mathbf{Z_2}$ on 
${\mathbb{R}}^1$, ${\sf{k}}_c$ becomes equal to zero. It is important to note that tension 
${V}_0^{\mathbb{H}}$ of the brane at the orbifold fixed point $z_0=0$ is half of what it would be without the orbifolding, {\it i.e.} 
${V}_0^{\mathbb{H}}={(2+N) \over \kappa^2} {\sf{k}}_0^{\mathbb{H}}$. The following argument
reveals this: Before orbifolding, the brane at the origin is split into two branes with 
equal tensions and located at $z=+\epsilon$ and $z=-\epsilon$. The $\mathbf{Z_2}$ 
symmetry is imposed and the equivalence class is moded out.
On taking the limit $\epsilon \rightarrow 0$, the brane at the fixed point
has half the original tension.  Note that the constraint now translates to
requiring that the sum of the brane tensions of this family must be positive.
The solution on the extra dimension
compactified on a circle is
\begin{eqnarray}
\sigma^{\mathbb{S}}(v) &=& \sum_{l=0}^2 {\sf{k}}_l^{\mathbb{S}} ~| v - v_l |
 + {\sf{k}}_c^{\mathbb{S}}  ~v ~ {\rm \ \ \  with \ \ \ }~
\sum_{l=0}^2 {\sf{k}}_l^{\mathbb{S}}\,v_l =  {\sf{k}}_c^{\mathbb{S}}\, \pi\, r ~ {\rm \ \ \  and\ \ \ }~\sum_{l=0}^2 {\sf{k}}_l^{\mathbb{S}} = 0 ~,
\label{S}
\end{eqnarray}
where the first constraint is just the continuity condition 
$\sigma^{\mathbb{S}}(0)=\sigma^{\mathbb{S}}(2 \pi r)$. The second condition states that the
sum of the brane tensions of this family must be zero. That this is required can be seen
by evaluating the sum of the jumps in the derivative 
$\frac{\partial \sigma^{\mathbb{S}}(v)}{\partial v}$ at each brane 
as one winds around the circle once. This sum is
$ \sum 2\,{\sf{k}}_l^{\mathbb{S}}$. The only value this quantity can take is zero since 
$\frac{\partial \sigma^{\mathbb{S}}(v)}{\partial v}$ is constant between each pair 
of branes and $\sigma^{\mathbb{S}}(v)$ does not depend on the winding number. For the orbifold compact dimension,
\begin{eqnarray}
\sigma^{\mathbb{I}}(w) &=& \sum_{{\rm{n-f.p.}}} {\sf{k}}_m^{\mathbb{I}} ~| w - w_m |
+ {\sf{k}}_t^{\mathbb{I}} ~w ~  {\rm \   with \  }~
{\sf{k}}_t^{\mathbb{I}} = \frac{1}{2}({\sf{k}}_0^{\mathbb{I}} - {\sf{k}}_{2}^{\mathbb{I}})~
\  {\rm and}~ \  {\sf{k}}_0^{\mathbb{I}}+{\sf{k}}_2^{\mathbb{I}}+\sum_{{\rm{n-f.p.}}} {2\,\sf{k}}_m^{\mathbb{I}} = 0 ~,
\label{orb}
\end{eqnarray}
where by ``n-f.p.'' we mean that the sum runs over all branes that are not located at the
two fixed points (or the sum runs over
``non-fixed point'' branes). Note that because $w_0$ and $w_2$ are fixed points, 
${\sf{k}}_0^{\mathbb{I}}={ \kappa^2  \over (2+N)} {V}_0^{\mathbb{I}} \,$ and
${\sf{k}}_2^{\mathbb{I}}={ \kappa^2  \over (2+N)} {V}_2^{\mathbb{I}} $. Thus, 
 the constraint in Eq.~(\ref{orb}) states that the sum of tensions of the branes constituting this family is zero. 

The functions in Eqs.~($\ref{R}-\ref{orb}$) must be consistent with 
Eq.~(\ref{dd0}) for them to be solutions to Einstein's equations. When 
$\Omega$ exists, the spacetime at the brane junctions is flat. Since 
Eq.~(\ref{dd0}) relates the brane tensions to the bulk cosmological constant, 
 the existence conditions serve as fine-tuning conditions that lead to a
Minkowski metric on the brane junctions.
To derive these, let us define 
\begin{eqnarray}
\chi_{j-1,j}^{\mathbb{R}} = \frac{\partial}{\partial y} \Omega^{-1} ~
\quad {\rm for} \quad y_{j-1} < y < y_{j},
\end{eqnarray}
with analogous definitions for $\chi_{k-1,k}^{\mathbb{H}}\,$, $\chi_{l-1,l}^{\mathbb{S}}\,$ and $\chi_{m-1,m}^{\mathbb{I}}\,$. In terms of the brane tensions, the bulk cosmological constant is
\begin{eqnarray}
\Lambda_{(j,k,l,m)} = - \frac{(2+N)(3+N)}{2\,\kappa^2}
~\{ (\chi_{j-1,j}^{\mathbb{R}})^2 + (\chi_{k-1,k}^{\mathbb{H}})^2
 + (\chi_{l-1,l}^{\mathbb{S}})^2 + (\chi_{m-1,m}^{\mathbb{I}})^2 \} ~.
\label{finetune}
\end{eqnarray}
It is worth emphasizing that even if the bulk cosmological constant is 
different in each cell, networks with orthogonally intersecting branes can be found.
\\

{{\bf Example Networks.}}
We will now use the machinery developed to study two networks:  
(i) The solution of Ref.~\cite{kaloper} will be shown to be a special case of our general formalism. This will serve as an example of how easy it is to 
generalize Eqs.~($\ref{R}-\ref{orb}$) to cases with an arbitrary number
of extra dimensions.
(ii) An illustration of a network with only positive tension branes will be provided.  

 Consider a network of $N$ orthogonal families of $(2+N)$-branes in a 
$(4+N)$-dimensional spacetime with a global bulk cosmological constant 
$\Lambda$. Assume that all the branes have the same magnitude of 
tension $V$, but that in
each family the tensions alternate in sign\footnote {In Ref.~\cite{kaloper}
the author starts with families containing branes with alternating tension and
assumes that the branes in different families have a different 
magnitude of tension. However, in the course of the calculation 
the author implicitly assumes that
the branes of all the families have the same magnitude of tension.}. 
Since we wish to reproduce the results of  Ref.~\cite{kaloper}, we take the
branes to be equidistant in each family.
Let us suppose that all the extra dimensions are $\mathbb{R}$. We will label
the extra dimensions by $y^i$, and the inter-brane separation in the dimension
$y^i$ by $l^i$. The number of branes in the family along $y^i$ is $2\,n^i+1$. 
Since $\Lambda$ is the same throughout the space, from Eqs.~(\ref{R}) and
(\ref{finetune}), we find ${\sf{k}}_c^{\mathbb{R}}=0$. Then,
\begin{eqnarray}
\Omega^{-1} &=& \sum_{i=1}^N\,\sigma^{\matbr}(y^i)+1\,, \\
\sigma^{\matbr}(y^i) &=& {\sf{k}}\,\sum_{j=0}^{2\,n^i}\,(-1)^{j}\,|y^i-y^i_j|-{\sf{k}}\,n^i\,l^i\,,
\label{kal}
\end{eqnarray}
where ${\sf{k}} = \frac{\kappa^2}{2\,(2+N)}\,V$  
and $y_j^i = j\,l^i$ denotes the positions of the branes in 
the direction $y^i$. Note that the brane network ends on positive 
tension branes. Equation~(\ref{kal}) can be rewritten as 
\begin{equation}
{\sigma^{\matbr}(y^i) \over {\sf{k}}} =2\,\sum_{j=0}^{2\,n^i}\,(-1)^{j}\,\theta(y^i-y^i_j)\,
(y^i-y^i_j)-y^i\,,
\end{equation}
which is Eq.~(10) of Ref.~\cite{kaloper} (where an accidental summation 
over $2\,n^i+2$ branes is performed when the number of branes in each 
family is $2\,n^i+1$).

We now present a network with only positive tension branes.
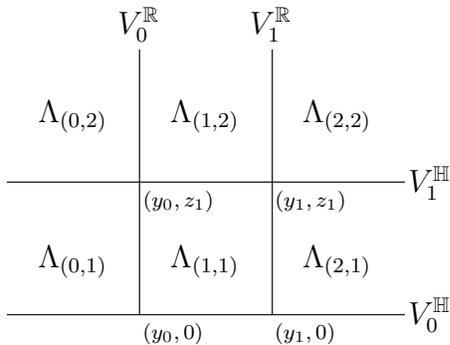
\begin{figure}[t]
%
\begin{center}
\begin{picture}(160,120)(0,0)

\Line(10,10)(160,10)
\Line(10,60)(160,60)
\Line(60,10)(60,110)
\Line(110,10)(110,110)
\Text(60,120)[]{$V_0^{\mathbb{R}}$}
\Text(110,120)[]{$V_1^{\mathbb{R}}$}
\Text(170,10)[]{$V_0^{\mathbb{H}}$}
\Text(170,60)[]{$V_1^{\mathbb{H}}$}
\Text(35,30)[]{$\Lambda_{(0,1)}$}
\Text(85,30)[]{$\Lambda_{(1,1)}$}
\Text(135,30)[]{$\Lambda_{(2,1)}$}
\Text(35,85)[]{$\Lambda_{(0,2)}$}
\Text(85,85)[]{$\Lambda_{(1,2)}$}
\Text(135,85)[]{$\Lambda_{(2,2)}$}
\Text(73,3)[]{\scriptsize{${(y_0,0)}$}}
\Text(123,3)[]{\scriptsize{$(y_1,0)$}}
\Text(75,53)[]{\scriptsize{$(y_0,z_1)$}}
\Text(125,53)[]{\scriptsize{$(y_1,z_1)$}}
\end{picture}
\end{center}
\caption[]{A network of four 4-branes in a 6-dimensional $AdS$ space. 
The $y$-dimension is ${\mathbb{R}}^1$ and the $z$-dimension is 
${\mathbb{R}}^1/{\mathbf{Z_2}}$.  All the branes
have positive tension.  }
\label{config}
\end{figure} 
From the constraints in Eqs.~($\ref{R}-\ref{orb}$), it can be noted that if
an extra dimension is either $\mathbb{S}$ or $\mathbb{I}$, the sum of the 
brane tensions in that family must be zero, 
and therefore it is necessary to include at least
one brane with negative tension. On the other hand, when the extra dimensions
are $\mathbb{R}$ or/and $\mathbb{H}$, the sum of the brane tensions is 
positive and it is possible to form networks with only positive tension branes.
As an example, consider a network of four 4-branes in a 6-dimensional 
 space where the $y$-dimension is ${\mathbb{R}}$ and the $z$-dimension is 
${\mathbb{H}}$ (see  Fig.~(\ref{config})).
 The conformal factor for this network is
\begin{eqnarray}
\Omega^{-1}(y,z) &=& {\sfk}_0^{\matbr}\,|y-y_0|
+ {\sfk}_1^{\matbr}\,|y-y_1|
+ {\sfk}_c^{\matbr}\,y+{\sfk}_1^{\matbh}\,|z-z_1|
+\,({\sfk}_0^{\matbh}+{\sfk}_1^{\matbh})\,z+\,c\,, \nonumber
\end{eqnarray}
where  $\kappa^2\,V_0^{\matbr} = 8\,{\sfk}_0^{\matbr}\ \,,\  
\kappa^2\,V_1^{\matbr} = 8\,{\sfk}_1^{\matbr}\ \,,\ 
\kappa^2\,V_0^{\matbh} = 4\,{\sfk}_0^{\matbh}\ $
and
$\,\kappa^2\,V_1^{\matbh} = 8\,{\sfk}_1^{\matbh}\,$. The fine-tuning
relations that make the metric on the brane junctions Minkowskian are
\begin{eqnarray}
\kappa^2\,\Lambda_{(0,1)} &=& 
-10\,\{({\sfk}_0^{\matbr}+{\sfk}_1^{\matbr}-{\sfk}_c^{\matbr})^2
+({\sfk}_0^{\matbh})^2 \} \,, \nonumber \\
\kappa^2\,\Lambda_{(1,1)} &=& 
-10\,\{({\sfk}_0^{\matbr}-{\sfk}_1^{\matbr}+{\sfk}_c^{\matbr})^2
+({\sfk}_0^{\matbh})^2 \} \,, \nonumber \\
\kappa^2\,\Lambda_{(2,1)} &=& 
-10\,\{({\sfk}_0^{\matbr}+{\sfk}_1^{\matbr}+{\sfk}_c^{\matbr})^2
+({\sfk}_0^{\matbh})^2 \} \,, \nonumber \\
\kappa^2\,\Lambda_{(0,2)} &=& 
-10\,\{({\sfk}_0^{\matbr}+{\sfk}_1^{\matbr}-{\sfk}_c^{\matbr})^2
+(2\,{\sfk}_1^{\matbh}+{\sfk}_0^{\matbh})^2 \} \,, \nonumber \\
\kappa^2\,\Lambda_{(1,2)} &=& 
-10\,\{({\sfk}_0^{\matbr}-{\sfk}_1^{\matbr}+{\sfk}_c^{\matbr})^2
+(2\, {\sfk}_1^{\matbh}+{\sfk}_0^{\matbh})^2 \} \,, \nonumber \\
\kappa^2\,\Lambda_{(2,2)} &=& 
-10\,\{({\sfk}_0^{\matbr}+{\sfk}_1^{\matbr}+{\sfk}_c^{\matbr})^2
+(2\,{\sfk}_1^{\matbh}+{\sfk}_0^{\matbh})^2 \} \,. \nonumber
\end{eqnarray}
\\

{{\bf Conclusion.}}
We have constructed the most general network of $N$ orthogonal
families of $(2+N)$-branes in $(4+N)$-dimensional $AdS$ spacetimes. 
We allowed the bulk cosmological constant to take an arbitrary value
in each cell formed by intersecting branes.  
The $(4+N)$-dimensional network is the direct product of the spaces 
defined by each extra dimension. Consequently, the network solution is
a linear combination of the solutions to Einstein's equations for
each extra dimension.
We presented results
for the cases in which the extra dimensions are the manifolds 
${\mathbb{R}}^1\,$, ${\mathbb{S}}^1$ and the 
orbifolds ${\mathbb{R}}^1/{\mathbf{Z}}_2\,$, 
${\mathbb{S}}^1/{\mathbf{Z}}_2$. Networks with only positive tension branes 
can be constructed if and only if the extra dimensions 
are ${\mathbb{R}}^1$ or/and 
${\mathbb{R}}^1/{\mathbf{Z}}_2\,$. For the purpose of illustration, we 
constructed two networks explicitly one of which consisted solely of positive
tension branes.
\\

{\it{Acknowledgments.}} 
This work was supported in part by a DOE grant No. DE-FG02-95ER40896 and 
in part by the Wisconsin Alumni Research Foundation.
\\
\\

\end{document}